\documentclass[12pt]{article}
\usepackage{graphicx}
\usepackage{color}
\usepackage{authblk}

\def\hybrid{\topmargin 0pt      \oddsidemargin 0pt
        \headheight 0pt \headsep 0pt
       \voffset-1cm
        \textwidth 6.25in       
       \textheight 9.5in       
        \marginparwidth 0.0in
        \parskip 5pt plus 1pt   \jot = 1.5ex}
\catcode`\@=11
\def\marginnote#1{}

\newcount\hour
\newcount\minute
\newtoks\amorpm
\hour=\time\divide\hour by60
\minute=\time{\multiply\hour by60 \global\advance\minute by-\hour}
\edef\standardtime{{\ifnum\hour<12 \global\amorpm={am}%
        \else\global\amorpm={pm}\advance\hour by-12 \fi
        \ifnum\hour=0 \hour=12 \fi
        \number\hour:\ifnum\minute<10 0\fi\number\minute\the\amorpm}}
\edef\militarytime{\number\hour:\ifnum\minute<10 0\fi\number\minute}

\def\draftlabel#1{{\@bsphack\if@filesw {\let\thepage\relax
   \xdef\@gtempa{\write\@auxout{\string
      \newlabel{#1}{{\@currentlabel}{\thepage}}}}}\@gtempa
   \if@nobreak \ifvmode\nobreak\fi\fi\fi\@esphack}
        \gdef\@eqnlabel{#1}}
\def\@eqnlabel{}
\def\@vacuum{}
\def\draftmarginnote#1{\marginpar{\raggedright\scriptsize\tt#1}}

\def\draftlabel#1{{\@bsphack\if@filesw {\let\thepage\relax
   \xdef\@gtempa{\write\@auxout{\string
      \newlabel{#1}{{\@currentlabel}{\thepage}}}}}\@gtempa
   \if@nobreak \ifvmode\nobreak\fi\fi\fi\@esphack}
        \gdef\@eqnlabel{#1}}
\def\@eqnlabel{}
\def\@vacuum{}
\def\draftmarginnote#1{\marginpar{\raggedright\scriptsize\tt#1}}

\def\draft{\oddsidemargin -.5truein
        \def\@oddfoot{\sl preliminary draft \hfil
        \rm\thepage\hfil\sl\today\quad\militarytime}
        \let\@evenfoot\@oddfoot \overfullrule 3pt
        \let\label=\draftlabel
        \let\marginnote=\draftmarginnote
   \def\@eqnnum{(\theequation)\rlap{\kern\marginparsep\tt\@eqnlabel}%
\global\let\@eqnlabel\@vacuum}  }

\def\numberbysection{\@addtoreset{equation}{section}
        \def\theequation{\thesection.\arabic{equation}}}

\def\underline#1{\relax\ifmmode\@@underline#1\else
        $\@@underline{\hbox{#1}}$\relax\fi}

\def\titlepage{\@restonecolfalse\if@twocolumn\@restonecoltrue\onecolumn
     \else \newpage \fi \thispagestyle{empty}\c@page\z@
        \def\thefootnote{\fnsymbol{footnote}} }

\def\endtitlepage{\if@restonecol\twocolumn \else  \fi
        \def\thefootnote{\arabic{footnote}}
        \setcounter{footnote}{0}}  
\relax

\numberbysection
\hybrid

\newfont{\Bbb}{msbm10 scaled 1\@ptsize00}
\newfont{\Bbbb}{msbm7 scaled 1\@ptsize00}

\newcommand{\DDD}{\raise-1pt\hbox{$\mbox{\Bbbb D}$}}

\newcommand{\UUU}{\raise-1pt\hbox{$\mbox{\Bbbb U}$}}

\newcommand{\z}{\raise-1pt\hbox{$\mbox{\Bbbb Z}$}}

\def\res{\mathop{\hbox{res}}\limits}

\def\beq{\begin{equation}}
\def\eeq{\end{equation}}
\def\p{\partial}

\begin{document}

\begin{titlepage}

\title{Matrix KP hierarchy and spin generalization of trigonometric
Calogero-Moser hierarchy}

\vspace{0.5cm}

\author[1,2]{V.~Prokofev\thanks{vadim.prokofev@phystech.edu }}
\author[3,4]{
 A.~Zabrodin\thanks{ zabrodin@itep.ru}}
 \affil[1]{Moscow Institute of Physics and Technology, Dolgoprudny, Institutsky per., 9,
Moscow region, 141700, Russia}
 \affil[2]{
Skolkovo Institute of Science and Technology, 143026 Moscow, Russian Federation
}
\affil[3]{ National Research University Higher School of Economics,
20 Myasnitskaya Ulitsa, Moscow 101000, Russian Federation}
\affil[4]{
Steklov Mathematical Institute of Russian Academy of Sciences,
Gubkina str. 8, Moscow, 119991, Russian Federation
}


\date{September 2019}
\maketitle

\vspace{-11cm} \centerline{ \hfill ITEP-TH-29/19}\vspace{11cm}

\begin{abstract}

We consider solutions of the matrix KP hierarchy that are trigonometric 
functions of the first hierarchical time $t_1=x$ and establish the correspondence
with the spin generalization of the trigonometric Calogero-Moser system on the level
of hierarchies. Namely, the evolution of poles $x_i$ and matrix residues at the poles 
$a_i^{\alpha}b_i^{\beta}$ of the solutions with respect to the 
$k$-th hierarchical time of the matrix KP hierarchy is shown to be given by the 
Hamiltonian flow with the 
Hamiltonian which is a linear combination of the first $k$ higher Hamiltonians of the 
spin 
trigonometric Calogero-Moser system with coordinates $x_i$ and with spin degrees of freedom
$a_i^{\alpha}, \, b_i^{\beta}$. By considering
evolution of poles according to the discrete time matrix KP hierarchy we also introduce 
the integrable discrete time version of the trigonometric spin Calogero-Moser system.

\end{abstract}

\end{titlepage}


\vspace{5mm}

\section{Introduction}

The matrix generalization of the Kadomtsev-Petviashvili (KP) hierarchy
is an infinite set of compatible nonlinear differential equations with
infinitely many independent (time) variables 
${\bf t}=\{t_1, t_2, t_3, \ldots \}$ and matrix dependent variables. 
It is a subhierarchy of the multi-component KP hierarchy
\cite{DJKM81,KL93,TT07,Teo11}. Among all solutions to these equations,
of special interest are solutions which have a finite number of poles 
in the variable $x=t_1$ in a fundamental domain of 
the complex plane. In particular, one can consider solutions
which are trigonometric or hyperbolic functions of $x$ with poles depending on the times
$t_2, t_3, \ldots$.

The dynamics of poles of singular solutions to nonlinear integrable
equations is a well known subject in mathematical physics 
\cite{AMM77,Krichever78,Krichever80,CC77}. 
It was shown
that the poles of solutions to the KP equation 
as functions of the time $t_2$ move 
as particles of the integrable Calogero-Moser many-body system
\cite{Calogero71,Calogero75,Moser75,OP81}. Rational, trigonometric and elliptic solutions
correspond respectively to rational, trigonometric or elliptic Calogero-Moser systems.

The further progress was achieved in \cite{Shiota94}, where it was shown that 
the correspondence between rational solutions to the KP equation and the Calogero-Moser
system with rational potential can be extended to the level of hierarchies. 
Namely, the evolution of poles with respect to the higher time $t_m$ 
of the KP hierarchy was shown to be given by 
the higher Hamiltonian flow  
of the integrable Calogero-Moser system with the Hamiltonian 
$H_m=\mbox{tr}\, L^m$, where $L$ is the Lax matrix. 
Later this correspondence was generalized to trigonometric
solutions of the KP hierarchy (see \cite{Haine07,Z19a}). It was shown
that the dynamics of poles with respect to the higher time $t_m$ 
is given by the Hamiltonian flow with the Hamiltonian
\beq\label{int1}
{\cal H}_m =\frac{1}{2(m+1)\gamma}\, \mbox{tr}\, \Bigl (
(L+\gamma I)^{m+1}- (L-\gamma I)^{m+1}\Bigr ),
\eeq
where $I$ is the unity matrix and $\gamma$ is a parameter 
such that $\pi i/\gamma$ is the period of the trigonometric or hyperbolic
functions. Clearly, the Hamiltonian ${\cal H}_m$ is a linear combination of the 
Hamiltonians $H_k=\mbox{tr}\, L^k$.

In this paper we generalize this result to trigonometric solutions of the matrix
KP hierarchy. The singular (in general, elliptic) solutions to the matrix KP
equation were investigated in \cite{KBBT95}.
It was shown that the evolution of data of such solutions
(positions of poles and some internal degrees of freedom)
with respect to the time $t_2$ is isomorphic to the dynamics of a spin generalization of the Calogero-Moser 
system (the Gibbons-Hermsen system \cite{GH84}). It is a system of ${\cal N}$
particles with coordinates $x_i$ with internal degrees of freedom 
given by $N$-dimensional column vectors ${\bf a}_i, {\bf b}_i$ which pairwise interact with each other.  
The Hamiltonian is
\beq\label{int2}
H=\sum_{i=1}^{{\cal N}}p_i^2-\gamma^2\sum_{i\neq k}
\frac{({\bf b}_i^T{\bf a}_k)({\bf b}_k^T{\bf a}_i)}{\sinh ^2(\gamma (x_i-x_k))}
\eeq
(here ${\bf b}_i^T$ is the transposed row-vector) with the non-vanishing Poisson brackets
$
\{x_i, p_k\}=\delta_{ik}, \, \{a_i^{\alpha}, b_k^{\beta}\}=
\delta_{\alpha \beta}\delta_{ik}
$.
The model is known to be integrable, with the higher Hamiltonians 
(integrals of motion in involution) 
$H_k=\mbox{tr} \, L^k$, where $L$ is the Lax matrix of the model given by
\beq\label{int3}
L_{jk}=-p_j \delta_{jk}- (1-\delta_{jk})\, 
\frac{\gamma \, {\bf b}_j^T {\bf a}_k}{\sinh (\gamma (x_j-x_k))}.
\eeq
Our main result in this paper is that 
the dynamics of poles $x_i$ and vectors ${\bf a}_i$, ${\bf b}_i$
(which parametrize matrix residues at the poles) 
with respect to the higher time $t_m$ 
is given by the Hamiltonian flow with the Hamiltonian 
(\ref{int1}) and with the Lax matrix (\ref{int3}). The corresponding result 
for rational solutions ($\gamma =0$) was established in \cite{PZ18}.

We use the method suggested by Krichever \cite{Krichever80} for elliptic solutions 
of the KP equation. It consists in substituting the solution not in the KP equation itself but
in the auxiliary linear problem for it (this implies a suitable pole ansatz for the wave
fuction). This method allows one to obtain the equations of motion together with the
Lax representation for them. 

Another result of this paper is the time discretization of the trigonometric
spin Calogero-Moser (Gibbons-Hermsen) model. (The time discretization of the 
rational spin Calogero-Moser system within the same approach 
was suggested in \cite{Z19aa}.)
Because of the precise correspondence between
the trigonometric solutions of the matrix KP hierarchy and the trigonometric 
spin Calogero-Moser hierarchy,
the integrable time discretization of the Calogero-Moser
system and its spin generalization can be obtained from dynamics of poles of 
trigonometric solutions to semi-discrete soliton equations. (``Semi'' means that 
the time becomes discrete while the space variable $x$, with respect to which
one considers pole solutions, remains continuous.)
At the same time, 
it is known that integrable discretizations of soliton equations can be
regarded as belonging to the same hierarchy as their continuous counterparts.
Namely, the discrete time step is equivalent to a special shift of infinitely many
continuous hierarchical times.
This fact lies in the basis of the method of generating discrete soliton
equations developed in \cite{DJM82}. For integrable time discretization of 
many-body systems see \cite{NP94,NRK96,RS97,Suris,KWZ98}.
In this paper, we derive equations of motion in discrete time $p$ for the
spin generalization of the trigonometric Calogero-Moser model: 
\beq\label{I2}
\begin{array}{c}
\displaystyle{
\sum_j 
\coth (\gamma (x_i(p)-x_j(p+1))
({\bf b}_i^T(p){\bf a}_j(p+1))({\bf b}^T_j(p+1)
{\bf a}_i(p))}
\\ \\
\displaystyle{
+\sum_j \coth (\gamma (x_i(p)-x_j(p-1))
({\bf b}^T_i(p){\bf a}_j(p-1))({\bf b}^T_j(p-1)
{\bf a}_i(p))}
\\ \\
\displaystyle{=
2\sum_{j\neq i} \coth (\gamma (x_i(p)-x_j(p))
({\bf b}^T_i(p){\bf a}_j(p))({\bf b}^T_j(p)
{\bf a}_i(p))},
\end{array}
\eeq
where ${\bf a}_i(p)$, ${\bf b}_i(p)$ are spin variables. 
In the limit $\gamma \to 0$ the result of \cite{Z19aa} is reproduced.

\section{The matrix KP hierarchy}

Here we briefly review the main facts about the multi-component and matrix
KP hierarchies following \cite{TT07,Teo11}.
We start from the more general multi-component KP hierarchy.
The independent variables are $N$ infinite sets of continuous ``times''
$$
{\bf t}=\{{\bf t}_1, {\bf t}_2, \ldots , {\bf t}_N\}, \qquad
{\bf t}_{\alpha}=\{t_{\alpha , 1}, t_{\alpha , 2}, t_{\alpha , 3}, \ldots \, \},
\qquad \alpha = 1, \ldots , N
$$
and $N$ discrete integer variables 
$
{\bf s}=\{s_1, s_2, \ldots , s_N\}
$
(``charges'') constrained by the condition
$\displaystyle{\sum_{\alpha =1}^N s_{\alpha}=0}$. In what follows, we will mostly
put $s_{\alpha}=0$ since we are interested in the dynamics in the continuous times.

In the bilinear formalism, 
the dependent variable is the tau-function $\tau ({\bf s};{\bf t})$. 
We also introduce the tau-functions
\beq\label{m6}
\tau _{\alpha \beta}({\bf t})=\tau ({\bf e}_{\alpha}-{\bf e}_{\beta}; {\bf t}),
\eeq
where
${\bf e}_{\alpha}$ is the vector whose $\alpha$th component is $1$ and all 
other entries are equal to zero.
The $N$-component KP hierarchy is the infinite set of bilinear equations
for the tau-functions which are encoded in the basic bilinear
relation
\beq\label{m5}
\sum_{\nu =1}^N \epsilon_{\alpha \nu}\epsilon_{\beta \nu}
\oint_{C_{\infty}}dz \, 
z^{\delta_{\alpha \nu}+\delta_{\beta \nu}-2}
e^{\xi ({\bf t}_{\nu}-{\bf t}_{\nu}', \, z)}
\tau _{\alpha \nu} \left ({\bf t}-[z^{-1}]_{\nu}\right )
\tau _{\nu \beta}\left ({\bf t}'+[z^{-1}]_{\nu}\right )=0
\eeq 
valid for any ${\bf t}$, ${\bf t}'$. Here $\epsilon_{\alpha \beta}$ is a sign factor:
$\epsilon_{\alpha \beta}=1$ if $\alpha \leq \beta$, $\epsilon_{\alpha \beta}=-1$
if $\alpha >\beta$.
In (\ref{m5}) we use the following standard
notation:
$$
\xi ({\bf t}_{\gamma}, z)=\sum_{k\geq 1}t_{\gamma , k}z^k,
$$
$$
\left ({\bf t}\pm [z^{-1}]_{\gamma}\right )_{\alpha k}=t_{\alpha , k}\pm
\delta_{\alpha \gamma} \frac{z^{-k}}{k}.
$$
The integration contour $C_{\infty}$ is a big circle around $\infty$.
Hereafter, we omit the variables ${\bf s}$ in the notation for 
the tau-functions.

An important role in the theory of integrable
hierarchies is played by the wave function. In the multi-component
KP hierarchy,
the wave function $\Psi ({\bf t};z)$ and its adjoint
$\Psi ^{\dag}({\bf t};z)$ are $N\! \times \! N$ matrices with the components
\beq\label{m2}
\begin{array}{l}
\displaystyle{\Psi_{\alpha \beta}({\bf t};z)=
\epsilon_{\alpha \beta}\,
\frac{\tau_{\alpha \beta} \left (
{\bf t}-[z^{-1}]_{\beta}\right )}{\tau ({\bf t})}\,
z^{\delta_{\alpha \beta}-1}e^{\xi ({\bf t}_{\beta}, z)},
}
\\ \\
\displaystyle{\Psi_{\alpha \beta}^{\dag}({\bf t};z)=
\epsilon_{\beta \alpha}\,
\frac{\tau _{\alpha \beta}\left (
{\bf t}+[z^{-1}]_{\alpha}\right )}{\tau ({\bf t})}\,
z^{\delta_{\alpha \beta}-1}e^{-\xi ({\bf t}_{\alpha}, z)}
}
\end{array}
\eeq
(here and below $\dag$ does not mean the Hermitian conjugation).
The complex variable $z$ is called the spectral parameter. 
Around $z=\infty$, the wave function $\Psi$ can be represented in 
the form of the series
\beq\label{m4}
\Psi_{\alpha \beta}({\bf t};z)=\left (\delta_{\alpha \beta}+
\sum_{k\geq 1}\frac{w^{(k)}_{\alpha \beta}( 
{\bf t})}{z^k}\right )e^{\xi ({\bf t}_{\beta}, z)},
\eeq
where $w^{(k)}({\bf t})$ are some matrix functions.
In terms of the wave functions, the bilinear relation (\ref{m5}) can be written
as
\beq\label{m3}
\oint_{C_{\infty}}\! dz \, \Psi ({\bf t};z)\Psi^{\dag} ({\bf t}';z)=0.
\eeq

Another (equivalent) approach to the multi-component KP hierarchy is based on matrix
pseudo-differential operators. 
The hierarchy can be understood as an infinite set of 
evolution equations in the times ${\bf t}$ for matrix functions of a variable $x$. 
For example, the coefficients $w^{(k)}$ of the wave function can be taken
as such matrix functions, the evolution being $w^{(k)}(x)\to w^{(k)}(x,{\bf t})$.
In what follows we denote $\tau (x, {\bf t})$, $w^{(k)}(x,{\bf t})$ simply as 
$\tau ({\bf t})$, $w^{(k)}({\bf t})$, suppressing the dependence on $x$. 
Let us introduce the matrix pseudo-differential ``wave operator'' ${\cal W}$ with matrix elements
\beq\label{m113}
{\cal W}_{\alpha \beta} = \delta_{\alpha \beta}+\sum_{k\geq 1}
w^{(k)}_{\alpha \beta}({\bf t})\p_x^{-k},
\eeq
where $w^{(k)}_{\alpha \beta}({\bf t})$ are the
same matrix functions as in (\ref{m4}).
The wave function 
is a result of action of the wave operator 
to the exponential function:
\beq\label{m113a}
\Psi ({\bf t}; z)={\cal W}
\exp \Bigl (xzI+\sum_{\alpha =1}^N E_{\alpha}\xi ({\bf t}_{\alpha}, z)\Bigr ),
\eeq
where $E_{\alpha}$ is the $N\! \times \! N$ matrix with the components
$(E_{\alpha})_{\beta \gamma}=\delta_{\alpha \beta}\delta_{\alpha \gamma}$. 
The adjoint wave function can be written as
\beq\label{m113b}
\Psi ^{\dag} ({\bf t}; z)=\exp \Bigl (-xzI-\sum_{\alpha =1}^N E_{\alpha}\xi ({\bf t}_{\alpha}, z)\Bigr )
{\cal W}^{-1}.
\eeq
Here the operators $\p_x$ 
which enter ${\cal W}^{-1}$ act to the left (the left action is defined as 
$f\p_x \equiv -\p_x f$).

It is proved in \cite{Teo11} that the wave function and its adjoint 
satisfy the linear
equations
\beq\label{m13c}
\p_{t_{\alpha , m}}\Psi ({\bf t}; z)=B_{\alpha m} \Psi ({\bf t}; z), 
\qquad
-\p_{t_{\alpha , m}}\Psi^{\dag} ({\bf t}; z)=\Psi^{\dag} ({\bf t}; z) B_{\alpha m} , 
\eeq
where $B_{\alpha m}$ is the differential operator
$
B_{\alpha m}= \Bigl ({\cal W} E_{\alpha}\p_x^m {\cal W}^{-1}\Bigr )_+.
$
The notation $(\ldots )_+$ means the differential part of a pseudo-differential operator, i.e.
the sum of all terms with $\p_x^k$, where $k\geq 0$. Again, the operator
$B_{\alpha m}$ in the second equation in (\ref{m13c}) acts to the left.
In particular, it follows from
(\ref{m13c}) at $m=1$ that
\beq\label{m13d}
\sum_{\alpha =1}^{N}\p_{t_{\alpha , 1}}\Psi ({\bf t}; z)=\p_x \Psi ({\bf t}; z),
\qquad
\sum_{\alpha =1}^{N}\p_{t_{\alpha , 1}}\Psi^{\dag} ({\bf t}; z)=\p_x \Psi^{\dag} ({\bf t}; z),
\eeq
so the vector field $\p_x$ can be identified with the vector field
$\sum_{\alpha }\p_{t_{\alpha , 1}}$.

The matrix KP hierarchy is a subhierarchy of the 
multi-component KP one which is obtained by
a restriction of the time 
variables in the following manner:
$t_{\alpha , m}=t_m$ for each $\alpha$ and $m$.
The corresponding vector fields are related as
$\p_{t_m}=\sum_{\alpha =1}^N \p_{t_{\alpha , m}}$.
The wave function for the matrix KP hierarchy has the expansion
\beq\label{m7}
\Psi_{\alpha \beta}({\bf t};z)=\left (\delta_{\alpha \beta}+
w_{\alpha \beta}^{(1)}({\bf t})z^{-1}+O(z^{-2})\right )
e^{xz+\xi ({\bf t}, z)},
\eeq
where $\displaystyle{\xi ({\bf t}, z)=\sum_{k\geq 1}t_kz^k}$.
The coefficient $w_{\alpha \beta}^{(1)}({\bf t})$ plays an important role in what follows.
Equations (\ref{m13c}) imply that the
wave function of the matrix KP hierarchy and its adjoint 
satisfy the linear
equations
\beq\label{m13a}
\p_{t_m}\Psi ({\bf t}; z)=B_m \Psi ({\bf t}; z), 
\qquad
-\p_{t_m}\Psi^{\dag} ({\bf t}; z)=\Psi^{\dag} ({\bf t}; z) B_m , \qquad m\geq 1,
\eeq
where $B_m$ is the differential operator
$
B_m= \Bigl ({\cal W} \p_x^m {\cal W}^{-1}\Bigr )_+.
$
At $m=1$ we have $\p_{t_1}\Psi =\p_x \Psi$, 
so we can identify
$\displaystyle{
\p_x = \p_{t_1}=\sum_{\alpha =1}^N \p_{t_{\alpha , 1}}}
$
and the evolution in $t_1$ is simply a shift of
the variable $x$:
$
w^{(k)}(x, t_1, t_2, \ldots )=w^{(k)}(x+t_1, t_2, \ldots ).
$
At $m=2$ equations (\ref{m13a}) turn into the linear problems
\beq\label{m14}
\p_{t_2}\Psi = \p_x^2\Psi +V({\bf t})\Psi ,  
\eeq
\beq\label{m14a}
-\p_{t_2}\Psi^{\dag} = \p_x^2\Psi^{\dag} +\Psi^{\dag}V({\bf t})
\eeq
which have the form of the matrix non-stationary Schr\"odinger equations with
\beq\label{m15}
V({\bf t})=-2\p_x w^{(1)}({\bf t}).
\eeq

Let us derive a useful corollary of the bilinear
relation (\ref{m5}).
Differentiating it with respect to $t_m$ and putting ${\bf t}'={\bf t}$ after this, we 
obtain:
\beq\label{m11}
\frac{1}{2\pi i}
\sum_{\nu =1}^N \oint_{C_{\infty}}dz \, z^m \Psi_{\alpha \nu}({\bf t}; z)
\Psi^{\dag}_{\nu \beta}({\bf t}; z)=-\p_{t_m}w_{\alpha \beta}^{(1)}({\bf t})
\eeq
or, equivalently, 
\beq\label{m11a}
\res_{\infty}\Bigl (z^m \Psi_{\alpha \nu}\Psi^{\dag}_{\nu \beta}
\Bigr ) =-\p_{t_m}w_{\alpha \beta}^{(1)}.
\eeq
Here and below the summation from $1$ to $N$ over repeated Greek indices is implied.
The residue at infinity is defined according to $\res_{\infty}\, (z^{-n})=\delta_{n1}$.

At the end of this section let us make some remarks on the discrete time version
of the matrix KP hierarchy.
The discrete time evolution is defined as a special shift 
of the infinite number of continuous time variables according to
the rule \cite{DJM82}
$$
\tau^p = \tau \left ({\bf t}-p\sum_{\alpha =1}^{N}[\mu^{-1}]_{\alpha}\right ), \qquad
\Psi^p=\Psi \left ({\bf t}-p\sum_{\alpha =1}^{N}[\mu^{-1}]_{\alpha};z\right ).
$$
Here $p$ is the discrete time variable and $\mu$ is a continuous parameter.
Each $\mu$ corresponds to its own discrete time flow. The limit $\mu \to \infty$ 
is the continuous limit. 
One can show, using the explicit expressions of the wave functions through the
tau-function and some corollaries of 
the bilinear relation (see \cite{Z19aa}) that the corresponding linear problems have the
form
\beq\label{l1a}
\mu \Psi_{\alpha \beta}^p -\mu \Psi_{\alpha \beta}^{p+1}=
\p_x \Psi_{\alpha \beta}^p+\left (w_{\alpha \nu}^{(1)}(p+1)-
w_{\alpha \nu}^{(1)}(p)\right )\Psi_{\nu \beta}^p,
\eeq
\beq\label{l2a}
\mu \Psi_{\alpha \beta}^{\dag p} -\mu \Psi_{\alpha \beta}^{\dag \, p-1}=
-\p_x \Psi_{\alpha \beta}^{\dag p}+
\Psi_{\alpha \nu}^{\dag p}\left (w_{\nu \beta}^{(1)}(p)-
w_{\nu \beta}^{(1)}(p-1)\right ).
\eeq

\section{Trigonometric solutions of the matrix KP hierarchy: dynamics of poles 
in $t_2$}

Our aim is to study solutions to the matrix KP hierarchy which are trigonometric functions
of the variable $x$ (and, therefore, $t_1$).
For the trigonometric solutions, the tau-function has the form
\beq\label{t1}
\tau = C\prod_{i=1}^{{\cal N}} (e^{2\gamma x}-e^{2\gamma x_i}),
\eeq
where $\gamma$ is a parameter. The period of the function is $\pi i/\gamma$.
Real (respectively, imaginary) $\gamma$ corresponds to hyperbolic (respectively, 
trigonometric) functions. In the limit $\gamma \to 0$ one obtains rational solutions. 
The ${\cal N}$ roots $x_i$ (assumed to be distinct) depend 
on the times ${\bf t}$. It is convenient to pass to the exponentiated variables
\beq\label{t2}
w=e^{2\gamma x}, \quad w_i =e^{2\gamma x_i},
\eeq
then the tau-function becomes a polynomial with the roots $w_i$:
$\tau = C \prod_i (w-w_i)$. Clearly, we have $\p_x = 2\gamma \p_w$,
$\p_x^2 = 4\gamma^2 (w^2 \p_w^2 + w\p_w )$. 

It is clear from (\ref{m2}) that the wave functions $\Psi$, $\Psi^\dag$
(and thus the coefficient $w^{(1)}$), 
as functions of $x$, 
have simple poles at $x=x_i$. It is shown in \cite{PZ18} that the residues 
at these poles are matrices of rank $1$. We parametrize them through the 
column vectors
${\bf a}_i =(a_i^1, a_i^2, \ldots , a_i^N)^T$,
${\bf b}_i =(b_i^1, b_i^2, \ldots , b_i^N)^T$,
${\bf c}_i=(c_i^1, c_i^2, \ldots , c_i^N)^T$ ($T$ means transposition)
and the row vector
${\bf c}^{*}_i=(c^{*1}_i, c^{*2}_i, \ldots , c^{*N}_i)$:
\beq\label{t3}
\Psi_{\alpha \beta}=e^{xz+\xi ({\bf t}, z)}\left (
C_{\alpha \beta}+\sum_i \frac{2\gamma w_i^{1/2}a_i^{\alpha}c_i^{\beta}}{w-w_i}\right ),
\eeq
\beq\label{t4}
\Psi^{\dag}_{\alpha \beta}=e^{-xz-\xi ({\bf t}, z)}\left (
C^{-1}_{\alpha \beta}+\sum_i \frac{2\gamma w_i^{1/2}
c_i^{*\alpha}b_i^{\beta}}{w-w_i}\right ),
\eeq
where the matrix $C_{\alpha \beta}$ does not depend on $x$. 
Note that
the constant term in the adjoint wave function is the inverse matrix
$C_{\alpha \beta}^{-1}$. This follows from (\ref{m113b}). 
For the matrices $w^{(1)}$ and $V=-2\p_x w^{(1)}$ we have
\beq\label{t5}
w^{(1)}_{\alpha \beta}=S_{\alpha \beta}-\sum_{i} 
\frac{2\gamma w_i a_i^{\alpha}b^{\beta}_i}{w-w_i},
\qquad
V_{\alpha \beta}=-8\gamma^2\sum_{i} \frac{ww_i a_i^{\alpha}b^{\beta}_i}{(w-w_i)^2},
\eeq
where the matrix $S_{\alpha \beta}$ does not depend on $x$. Tending $w\to \infty$ in (\ref{m11a}),
one concludes that $\p_{t_m}S_{\alpha \beta}=0$ for all 
$m\geq 1$, so the matrix $S_{\alpha \beta}$ does not depend on 
all the times. The components of the vectors ${\bf a}_i$, ${\bf b}_i$ are going to be
spin variables of the Gibbons-Hermsen model. 

We first consider the dynamics of poles 
with respect to the time $t_2$. The procedure is similar to the rational 
case \cite{PZ18}. Following Krichever's approach, 
we consider the linear problems
(\ref{m14}), (\ref{m14a}),
$$
\p_{t_2}\Psi_{\alpha \beta}=\p_{x}^2 \Psi_{\alpha \beta}-8\gamma^2
\sum_{i=1}^{{\cal N}} \frac{ww_ia_i^{\alpha}b^{\nu}_i}{(w-w_i)^2}\, \Psi_{\nu \beta},
$$
$$
-\p_{t_2}\Psi_{\alpha \beta}^{\dag}=\p_{x}^2 \Psi_{\alpha \beta}^{\dag}-8\gamma^2
\Psi^{\dag}_{\alpha \nu} \sum_{i=1}^{{\cal N}}\frac{ww_ia_i^{\nu}b^{\beta}_i}{(w-w_i)^2}
$$
and substitute here the pole ansatz (\ref{t3}), (\ref{t4}) for the wave functions. 
Consider first the equation for $\Psi$. 
First of all, comparing the behavior of both sides as $w\to \infty$, we conclude that
$\p_{t_2}C_{\alpha \beta}=0$, so $C_{\alpha \beta}$ does not depend on $t_2$
(in a similar way, from the higher linear problems one can see that 
$C_{\alpha \beta}$ does not depend on all the times 
$t_m$). After the substitution, we see that the expression has poles at $w=w_i$
up to the third order. 
Equating coefficients at the poles of different orders at $w=w_i$,
we get the conditions:
\begin{itemize}
\item
At $\frac{1}{(w-w_i)^3}$: \phantom{a} $b_i^{\nu}a_i^{\nu}=1$;
\item
At $\frac{1}{(w-w_i)^2}$: \phantom{a} $\displaystyle{
-\frac{1}{2}\, \dot x_i c_i^{\beta}-2\gamma \sum_{k\neq i}
\frac{w_i^{1/2}w_k^{1/2}b_i^{\nu}a_k^{\nu}c_k^{\beta}}{w_i-w_k}-
(z-\gamma )c_i^{\beta}=w_i^{1/2}\tilde b_i^{\beta}}$;
\item
At $\frac{1}{w-w_i}$: $$\p_{t_2}(w_i^{1/2}
a_i^{\alpha}c_i^{\beta})=2\gamma w_i^{1/2}\dot x_i a_i^{\alpha}c_i^{\beta}
+8\gamma^2
\sum_{k\neq i}\frac{w_i^2 w_k^{1/2}
a_i^{\alpha}b_i^{\nu}a_k^{\nu}c_k^{\beta}}{(w_i-w_k)^2}-8\gamma^2
\sum_{k\neq i}\frac{w_i^{3/2}w_k
a_k^{\alpha}b_k^{\nu}a_i^{\nu}c_i^{\beta}}{(w_i-w_k)^2},
$$
\end{itemize}
where $\tilde b_i^{\beta}=b_i^{\nu}C_{\nu \beta}$,
and $\dot x_i=\p_{t_2}x_i$. 
Similar calculations for the linear problem for $\Psi^{\dag}$ lead to the conditions
\begin{itemize}
\item
At $\frac{1}{(w-w_i)^3}$: \phantom{a} $b_i^{\nu}a_i^{\nu}=1$;
\item
At $\frac{1}{(w-w_i)^2}$: \phantom{a} $\displaystyle{
-\frac{1}{2}\, \dot x_i c_i^{*\alpha}-2\gamma \sum_{k\neq i}
\frac{w_i^{1/2}w_k^{1/2}c_k^{*\alpha}b_k^{\nu}a_i^{\nu}}{w_k-w_i}-
(z+\gamma )c_i^{*\alpha}=-w_i^{1/2}\tilde a_i^{\alpha}}$;
\item
At $\frac{1}{w-w_i}$:
\end{itemize}
$$\p_{t_2}(w_i^{1/2}
c_i^{*\alpha}b_i^{\beta})=-2\gamma w_i^{1/2}\dot x_i c_i^{*\alpha}b_i^{\beta}
+8\gamma^2
\sum_{k\neq i}\frac{w_i^2 w_k^{1/2}
c_k^{*\alpha}a_i^{\nu}b_k^{\nu}b_i^{\beta}}{(w_i-w_k)^2}-8\gamma^2
\sum_{k\neq i}\frac{w_i^{3/2}w_k
c_i^{*\alpha}a_k^{\nu}b_i^{\nu}b_k^{\beta}}{(w_i-w_k)^2},
$$
where $\tilde a_i^{\alpha}=C^{-1}_{\alpha \nu}a_i^{\nu}$.

The conditions coming from the
third order poles are constraints on the vectors ${\bf a}_i$, ${\bf b}_i$. 
The other conditions can be written in the matrix form
\beq\label{t6}
\left \{
\begin{array}{l}
(zI-(L+\gamma I)){\bf c}^{\beta}=-W^{1/2}\tilde {\bf b}^{\beta}
\\ \\
\dot {\bf c}^{\beta}=M{\bf c}^{\beta},
\end{array}
\right.
\eeq
\beq\label{t7}
\left \{
\begin{array}{l}
{\bf c}^{*\alpha}(zI-(L-\gamma I))=\tilde {\bf a}^{\alpha T}W^{1/2}
\\ \\
\dot {\bf c}^{*\alpha}={\bf c}^{*\alpha} \tilde M,
\end{array}
\right.
\eeq
where ${\bf c}^{\beta}=(c^{\beta}_1, \ldots c^{\beta}_{\cal N})^T$,
${\bf c}^{*\alpha}=(c^{*\alpha}_1, \ldots c^{*\alpha}_{\cal N})$,
$\tilde {\bf b}^{\beta}=(\tilde b^{\beta}_1, \ldots \tilde b^{\beta}_{\cal N})^T$,
$\tilde {\bf a}^{\alpha}=(\tilde a^{\alpha}_1, \ldots \tilde a^{\alpha}_{\cal N})$
are ${\cal N}$-dimensional vectors, $I$ is the unity matrix, 
$W=\mbox{diag}\, (w_1, w_2, \ldots w_{\cal N})$ and $L$, $M$, $\tilde M$
are ${\cal N}\! \times \! {\cal N}$ matrices of the form
\beq\label{t8}
L_{ik}=-\frac{1}{2}\, \dot x_i \delta_{ik}-2\gamma (1-\delta_{ik})\,
\frac{w_i^{1/2}w_k^{1/2}b_i^{\nu}a_k^{\nu}}{w_i-w_k},
\eeq
\beq\label{t9}
M_{ik}=(\gamma \dot x_i -\Lambda_i)\delta_{ik}+8\gamma^2 (1-\delta_{ik})\,
\frac{w_i^{3/2}w_k^{1/2}b_i^{\nu}a_k^{\nu}}{(w_i-w_k)^2},
\eeq
\beq\label{t10}
\tilde M_{ik}=(\gamma \dot x_i +\Lambda_i^*)\delta_{ik}-8\gamma^2 (1-\delta_{ik})\,
\frac{w_i^{1/2}w_k^{3/2}b_i^{\nu}a_k^{\nu}}{(w_i-w_k)^2}.
\eeq
Here
\beq\label{t11}
\Lambda_i=\frac{\dot a_i^{\alpha}}{a_i^{\alpha}}+8\gamma^2
\sum_{k\neq i}\frac{w_iw_k a_k^{\alpha}b_k^{\nu}a_i^{\nu}}{a_i^{\alpha}(w_i-w_k)^2},
\quad
-\Lambda_i^*=\frac{\dot b_i^{\alpha}}{b_i^{\alpha}}-8\gamma^2
\sum_{k\neq i}\frac{w_iw_k b_i^{\nu}a_k^{\nu}b_k^{\alpha}}{b_i^{\alpha}(w_i-w_k)^2}
\eeq
do not depend on the index $\alpha$. In fact one can see that $\Lambda_i=\Lambda_i^*$.
Indeed, multiplying equations (\ref{t11}) by $a_i^{\alpha}b_i^{\alpha}$
(no summation here!), summing over 
$\alpha$ and summing the two equations, we get $\Lambda_i-\Lambda_i^*=\p_{t_2}(
a_i^{\alpha}b_i^{\alpha})=0$ by virtue of the constraint
$a_i^{\alpha}b_i^{\alpha}=1$. 

Differentiating the first equation in (\ref{t6}) by $t_2$, we get, after some
calculations, the compatibility condition of equations (\ref{t6}):
\beq\label{t12}
(\dot L+[L,M]){\bf c}^{\beta}=0.
\eeq
One can see, taking into account equations (\ref{t11}), which we write here in the form
\beq\label{t13}
\dot a_i^{\alpha}=\Lambda_i a_i^{\alpha}-2\gamma^2
\sum_{k\neq i}\frac{a_k^{\alpha}b_k^{\nu}a_i^{\nu}}{\sinh^2(\gamma (x_i-x_k))},
\eeq
\beq\label{t13a}
\dot b_i^{\alpha}=-\Lambda_i b_i^{\alpha}+2\gamma^2
\sum_{k\neq i}\frac{b_i^{\nu}a_k^{\nu}b_k^{\alpha}}{\sinh^2(\gamma (x_i-x_k))}
\eeq
(in this form they are equations of motion for the spin degrees of freedom)
that the off-diagonal elements of the matrix $\dot L+[L,M]$ are equal to zero.
Vanishing of the diagonal elements yields equations of motion for the poles $x_i$:
\beq\label{t14}
\ddot x_i=-8\gamma^3 \sum_{k\neq i}
\frac{\cosh (\gamma (x_i-x_k))}{\sinh^3 (\gamma (x_i-x_k))}\, b_i^{\mu}a_k^{\mu}
b_k^{\nu}a_i^{\nu}.
\eeq
The gauge transformation $a_i^{\alpha}\to a_i^{\alpha}q_i$,
$b_i^{\alpha}\to b_i^{\alpha}q_i^{-1}$ with
$\displaystyle{q_i=\exp \Bigl (\int^{t_2}\Lambda_i dt\Bigr )}$ eliminates 
the terms with $\Lambda_i$ in (\ref{t13}), (\ref{t13a}), so we can put
$\Lambda_i=0$. This gives the equations of motion
\beq\label{t15}
\dot a_i^{\alpha}=-2\gamma^2
\sum_{k\neq i}\frac{a_k^{\alpha}b_k^{\nu}a_i^{\nu}}{\sinh^2(\gamma (x_i-x_k))},
\quad
\dot b_i^{\alpha}=2\gamma^2
\sum_{k\neq i}\frac{b_i^{\nu}a_k^{\nu}b_k^{\alpha}}{\sinh^2(\gamma (x_i-x_k))}.
\eeq
Together with (\ref{t14}) they are equations of motion of the trigonometric 
Gibbons-Hermsen model.
Their Lax representation is given by the matrix equation $\dot L=[M,L]$. It states that
the time evolution of the Lax matrix is an isospectral transformation. It follows 
that the quantities $H_k=\mbox{tr}\, L^k$ are integrals of motion. In particular,
\beq\label{t16}
H_2=\sum_{i=1}^{{\cal N}}p_i^2-\gamma^2\sum_{i\neq k}
\frac{b_i^{\mu}a_k^{\mu} b_k^{\nu}a_i^{\nu}}{\sinh ^2(\gamma (x_i-x_k))}=
\mbox{tr}\, L^2
\eeq
is the Hamiltonian of the Gibbons-Hermsen model. Equations of motion (\ref{t15}),
(\ref{t14}) are equivalent to the Hamiltonian equations
\beq\label{t17}
\dot x_i=\frac{\p H_2}{\p p_i}, \quad \dot p_i=-\frac{\p H_2}{\p x_i},
\quad
\dot a_i^{\alpha}=\frac{\p H_2}{\p b_i^{\alpha}}, \quad
\dot b_i^{\alpha}=-\frac{\p H_2}{\p a_i^{\alpha}}.
\eeq

\section{Dynamics of poles in the higher times}

The main tool for the analysis of the dynamics in the higher times 
is the relation (\ref{m11a}) which, after substitution of (\ref{t3}),
(\ref{t4}) and (\ref{t5}) takes the form
\beq\label{d0}
\begin{array}{c}
\displaystyle{\res_{\infty} \left [z^m \Bigl (C_{\alpha \nu}+\sum_i
\frac{2\gamma w_i^{1/2}a_i^{\alpha}c_i^{\nu}}{w-w_i}\Bigr )
\Bigl (C^{-1}_{\nu \beta}+\sum_k \frac{2\gamma w_k^{1/2}c_k^{*\nu}b_k^{\beta}}{w-w_k}
\Bigr )\right ]}
\\ \\
\displaystyle{
=\, 2\gamma \sum_i \frac{\p_{t_m}(w_i a_i^{\alpha}b_i^{\beta})}{w-w_i}+
4\gamma^2 \sum_i \frac{\p_{t_m}x_i\, w_i^2 a_i^{\alpha}b_i^{\beta}}{(w-w_i)^2}.}
\end{array}
\eeq
The both sides are rational functions of $w$ with poles at
$w=w_i$ vanishing at infinity. Identifying the coefficients in front of the 
second order poles, we obtain
\beq\label{d1}
\p_{t_m}x_i=\res_{\infty}\Bigl (z^m c_i^{\nu}w_i^{-1}c_i^{*\nu}\Bigr ).
\eeq
Solving the linear equations (\ref{t6}), (\ref{t7}), we get
\beq\label{d1a}
c_i^{\nu}=-\sum_k (zI-(L+\gamma I))^{-1}_{ik}w_k^{1/2}\tilde b_k^{\nu},
\quad
c_i^{*\nu}=\sum_k \tilde a_k^{\nu}w_k^{1/2}(zI-(L-\gamma I))^{-1}_{ki},
\eeq
and, therefore, (\ref{d1}) reads
$$
\p_{t_m}x_i=-\res_{\infty}\sum_{k,k'}\left (z^m \tilde a_k^{\nu}
\tilde b_{k'}^{\nu}w_k^{1/2}\Bigl (\frac{1}{zI-(L-\gamma I)}\Bigr )_{ki}
w_i^{-1} \Bigl (\frac{1}{zI-(L+\gamma I)}\Bigr )_{ik'}w_{k'}^{1/2}\right )
$$
$$
=-\res_{\infty}\, \mbox{tr} \left (z^m W^{1/2}RW^{1/2}
\frac{1}{zI-(L-\gamma I)}\, W^{-1}E_i \frac{1}{zI-(L+\gamma I)}\right ),
$$
where $E_i$ is the diagonal matrix with matrix elements $(E_i)_{jk}=\delta_{ij}\delta_{ik}$
and $R$ is the ${\cal N}\! \times \! {\cal N}$ matrix
\beq\label{d2}
R_{ik}=\tilde b_i^{\nu}\tilde a_k^{\nu}=b_i^{\nu} a_k^{\nu}.
\eeq
The following commutation relation can be checked directly:
\beq\label{d3}
[L, W]=2\gamma (W^{1/2}RW^{1/2}-W).
\eeq
Note that $E_i=-\p L/ \p p_i$. 
The rest of the calculation is similar to the one done in \cite{Z19a}. 
We have, using (\ref{d3}):
$$
\p_{t_m}x_i=\frac{1}{2\gamma}\res_{\infty}\, \mbox{tr}
\left (z^m (LW-WL +2\gamma W)\, 
\frac{1}{zI-(L-\gamma I)}\, W^{-1}\frac{\p L}{\p p_i} \frac{1}{zI-(L+\gamma I)}\right )
$$
$$
=\frac{1}{2\gamma}\res_{\infty}\, \mbox{tr}\left (
z^m \left (\frac{\p L}{\p p_i}\frac{1}{zI-(L+\gamma I)}-
\frac{\p L}{\p p_i}\frac{1}{zI-(L-\gamma I)}\right ) \right )
$$
$$
=\frac{1}{2\gamma}\, \mbox{tr}\left (\frac{\p L}{\p p_i}(L+\gamma I)^m -
\frac{\p L}{\p p_i}(L-\gamma I)^m \right )
$$
$$
=\frac{1}{2(m+1)\gamma}\, 
\frac{\p }{\p p_i}\, \mbox{tr}\Bigl ((L+\gamma I)^{m+1}-(L-\gamma I)^{m+1}\Bigr )
=\frac{\p {\cal H}_m}{\p p_i},
$$
where ${\cal H}_m$ is given by (\ref{int1}). Note that
${\cal H}_2 =H_2 +\mbox{const}$. We have obtained one part of the 
Hamiltonian equations for the higher time flows. In the case $\gamma \to 0$ 
(rational solutions) the result of the paper \cite{PZ18} is reproduced. 

In order to obtain another part of the Hamiltonian equations, let us differentiate
(\ref{d1}) with respect to $t_2$: 
$$
\p_{t_m}\dot x_i= -2\gamma \res_{\infty}\Bigl (z^m c_i^{*\nu} \dot x_i w_i^{-1}
c_i^{\nu} \Bigr ) +\res_{\infty} \Bigl (z^m (c_i^{\nu}  w_i^{-1}
\p_{t_2}c_i^{\nu} + \p_{t_2}c_i^{* \nu} w_i^{-1}
c_i^{\nu} )\Bigr )
$$
$$
=\res_{\infty}\sum_k \Bigl (z^m (c_i^{*\nu}w_i^{-1}B_{ik}c_k^{\nu}-
c_k^{*\nu}w_k^{-1}B_{ki}c_i^{\nu})\Bigr ),
$$
where
$$
B_{jk}=8\gamma^2 (1-\delta_{jk})\, \frac{w_j^{3/2}w_k^{1/2}
b_j^{\nu}a_k^{\nu}}{(w_i-w_k)^2}.
$$
Therefore, we have, using (\ref{d1a}):
$$
\p_{t_m}p_i = \frac{1}{2}\, \p_{t_m}\dot x_i 
=-\res_{\infty}\left [ z^m \, \mbox{tr} \left ( W^{1/2}RW^{1/2}\frac{1}{zI-(L-\gamma I)}
\, G^{(i)} \frac{1}{zI-(L+\gamma I)}\right )\right ],
$$
where the matrix $G^{(i)}$ is given by
$$
G^{(i)}_{jk}=4\gamma^2 (\delta_{ij}-\delta_{ik})\,
\frac{w_j^{1/2}w_k^{1/2}
b_j^{\nu}a_k^{\nu}}{(w_i-w_k)^2}.
$$
It is straightforward to check the identities
$$
(WG^{(i)}\! -\! G^{(i)}W)_{jk}=-2\gamma L_{jk}(\delta_{ij}-\delta_{ik}),
$$
$$
WG^{(i)}\! +\! G^{(i)}W =2 \frac{\p L}{\p x_i}.
$$
A direct calculation which literally repeats the one done in \cite{Z19a}
shows that
$$
\p_{t_m}p_i = 
=-\frac{1}{2\gamma}
\, \res_{\infty}\left [ z^m \, \mbox{tr} \left ( 
(LW-WL+2\gamma W)\frac{1}{zI-(L-\gamma I)}
\, G^{(i)} \frac{1}{zI-(L+\gamma I)}\right )\right ]
$$
$$
=-\frac{1}{2\gamma}\res_{\infty}\left [z^m \mbox{tr} 
\left (WG^{(i)} \frac{1}{zI-(L+\gamma I)}- G^{(i)}W \frac{1}{zI-(L-\gamma I)}\right )\right ]
$$
$$
=-\frac{1}{2\gamma}\res_{\infty}\left [z^m \mbox{tr}
\left (\frac{\p L}{\p x_i}
\left (\frac{1}{zI-(L+\gamma I)}-\frac{1}{zI-(L-\gamma I)}\right )\right ) \right ]
$$
$$
=-\frac{1}{2\gamma}\, \mbox{tr} \left (\frac{\p L}{\p x_i}\,
(L+\gamma I)^m -\frac{\p L}{\p x_i} \, (L-\gamma I)^m \right )
=-\, \frac{\p {\cal H}_m}{\p x_i}.
$$
We have established the remaining part of the Hamiltonian equations for the 
higher time dynamics
of the $x_i$'s. 

\section{Dynamics of spin variables in the higher times}

Comparison of the first order poles in (\ref{d0}) gives the following relation:
$$
\p_{t_m}(w_i a_i^{\alpha}b_i^{\beta})=\res_{\infty} \left [
z^m \Bigl (w_i^{1/2}C_{\alpha \nu}c_i^{*\nu}b_i^{\beta}+w_i^{1/2}
a_i^{\alpha}c_i^{\nu}C_{\nu \beta}^{-1}\phantom{\sum_{k\neq i}}
\right.
$$
$$
\left. 
+2\gamma \sum_{k\neq i}\frac{w_i^{1/2}w_k^{1/2}}{w_i-w_k}\,
(a_i^{\alpha}b_k^{\beta}c_i^{\nu}c_k^{*\nu}+a_k^{\alpha}b_i^{\beta}c_k^{\nu}c_i^{*\nu})
\Bigr )\right ].
$$
Using (\ref{d1a}), we can rewrite it in the form
$$
b_i^{\beta}\left [ -\p_{t_m}a_i^{\alpha}+\res_{\infty} \Bigl (z^m \Bigl (
\sum_k a_k^{\alpha}w_i^{-1/2}w_k^{1/2}\Bigl (\frac{1}{zI-(L-\gamma I)}\Bigr )_{ki}
\right.
$$
$$
\left.
-2\gamma \sum_{k\neq i}\sum_{l,n}\frac{w_i^{-1/2}w_k^{1/2}}{w_i-w_k}\,
a_k^{\alpha}a_l^{\nu}w_l^{1/2}\Bigl (\frac{1}{zI-(L-\gamma I)}\Bigr )_{li}
\Bigl (\frac{1}{zI-(L+\gamma I)}\Bigr )_{kn}w_n^{1/2}b_n^{\nu}\Bigr ) \Bigr )\right ]
$$
$$
-a_i^{\alpha}\left [ \p_{t_m}b_i^{\beta}+\res_{\infty} \Bigl (z^m \Bigl (
\sum_k b_k^{\beta}w_i^{-1/2}w_k^{1/2}\Bigl (\frac{1}{zI-(L+\gamma I)}\Bigr )_{ik}
\right.
$$
$$
\left.
+2\gamma \sum_{k\neq i}\sum_{l,n}\frac{w_i^{-1/2}w_k^{1/2}}{w_i-w_k}\,
b_k^{\beta}a_l^{\nu}w_l^{1/2}\Bigl (\frac{1}{zI-(L-\gamma I)}\Bigr )_{lk}
\Bigl (\frac{1}{zI-(L+\gamma I)}\Bigr )_{in}w_n^{1/2}b_n^{\nu}\Bigr ) \Bigr )\right ]
$$
$$
=2\gamma \p_{t_m}x_i \, a_i^{\alpha}b_i^{\beta}.
$$
Separating the terms at $k=i$ in the sums over $k$ in the first and the third lines,
and taking into account that
$$
2\gamma \p_{t_m}x_i = \res_{\infty}\, \mbox{tr}\left [z^m E_i \Bigl (
\frac{1}{zI-(L-\gamma I)}-\frac{1}{zI-(L+\gamma I)}\Bigr )\right ]
$$
$$
=\res_{\infty}\left [z^m \Bigl (
\frac{1}{zI-(L-\gamma I)}\Bigr )_{ii}-z^m \Bigl (\frac{1}{zI-(L+\gamma I)}\Bigr )_{ii}
\right ],
$$
we represent this equation as follows: 
\beq\label{d4a}
b_i^{\beta}P_i^{\alpha}-a_i^{\alpha}Q_i^{\beta}=0,
\eeq
where
$$
P_i^{\alpha}=-\p_{t_m}a_i^{\alpha}+\res_{\infty} \left [z^m \left (
\sum_{k\neq i} a_k^{\alpha}w_i^{-1/2}w_k^{1/2}\Bigl (\frac{1}{zI-(L-\gamma I)}\Bigr )_{ki}
\right. \right.
$$
$$
\left. \left.
+\mbox{tr}\left (W^{1/2}RW^{1/2}\frac{1}{zI-(L-\gamma I)}W^{-1}
\frac{\p L}{\p b_i^{\alpha}}\, \frac{1}{zI-(L+\gamma I)}\right ) \right )\right ],
$$
$$
Q_i^{\beta}=\p_{t_m}b_i^{\beta}+\res_{\infty} \left [z^m \left (
\sum_{k\neq i} b_k^{\beta}w_i^{-1/2}w_k^{1/2}\Bigl (\frac{1}{zI-(L+\gamma I)}\Bigr )_{ik}
\right. \right.
$$
$$
\left. \left.
+\mbox{tr}\left (W^{1/2}RW^{1/2}\frac{1}{zI-(L-\gamma I)}
\frac{\p L}{\p a_i^{\beta}}\, W^{-1} \frac{1}{zI-(L+\gamma I)}\right ) \right )\right ],
$$
Here we took into account that 
$$
\frac{\p L_{jk}}{\p b_i^{\alpha}}=-2\gamma \delta_{ij}(1-\delta_{jk})\,
\frac{w_i^{1/2}w_k^{1/2}a_k^{\alpha}}{w_i-w_k}, \quad
\frac{\p L_{jk}}{\p a_i^{\alpha}}=-2\gamma \delta_{ik}(1-\delta_{jk})\,
\frac{w_j^{1/2}w_i^{1/2}b_j^{\alpha}}{w_j-w_i}.
$$
It then follows from (\ref{d4a}) that
\beq\label{d4}
\frac{P_i^{\alpha}}{a_i^{\alpha}}=\frac{Q_i^{\beta}}{b_i^{\beta}}=-\Lambda ^{(m)}_i
\eeq
does not depend on the indices $\alpha$, $\beta$. 

Let us transform the expressions for $P_i^{\alpha}$, $Q_i^{\beta}$ using the 
commutation relation (\ref{d3}), i.e., substituting
$$
W^{1/2}RW^{1/2}=\frac{1}{2\gamma}\, (LW-WL +2\gamma W).
$$
We have:
$$
P_i^{\alpha}=-\p_{t_m}a_i^{\alpha}+\frac{1}{2\gamma}
\res_{\infty} \left [z^m \left (
\mbox{tr}\left (\frac{\p L}{\p b_i^{\alpha}}\,
\frac{1}{zI-(L+\gamma I)}-
\frac{\p L}{\p b_i^{\alpha}}\,
\frac{1}{zI-(L-\gamma I)}\right )\right.\right.
$$
$$
\left. \left.
+2\gamma \sum_{k\neq i} a_k^{\alpha}w_i^{-1/2}w_k^{1/2}
\Bigl (\frac{1}{zI-(L-\gamma I)}\Bigr )_{ki} +
\mbox{tr}\Bigl (\frac{\p L}{\p b_i^{\alpha}}-W^{-1}\frac{\p L}{\p b_i^{\alpha}}\, W\Bigr )
\frac{1}{zI-(L-\gamma I)}\right )\right ].
$$
But
$$
\left (\frac{\p L}{\p b_i^{\alpha}}-W^{-1}\frac{\p L}{\p b_i^{\alpha}}\, W\right )_{jk}=
-2\gamma \delta_{ij}(1-\delta_{jk})w_i^{-1/2}w_k^{1/2}a_k^{\alpha},
$$
ans so the second line is equal to zero. We are left with
\beq\label{d5}
P_i^{\alpha}=-\p_{t_m}a_i^{\alpha}+\frac{\p {\cal H}_m}{\p b_i^{\alpha}}.
\eeq
A similar calculation for $Q_i^{\alpha}$ yields
\beq\label{d5a}
Q_i^{\alpha}=\p_{t_m}b_i^{\alpha}+\frac{\p {\cal H}_m}{\p a_i^{\alpha}}.
\eeq
Therefore, from (\ref{d4}) we have the equations of motion
$$
\p_{t_m}a_i^{\alpha}=\frac{\p {\cal H}_m}{\p b_i^{\alpha}}+
\Lambda_i^{(m)}a_i^{\alpha},
\quad
\p_{t_m}b_i^{\alpha}=-\frac{\p {\cal H}_m}{\p a_i^{\alpha}}-
\Lambda_i^{(m)}b_i^{\alpha}.
$$
The gauge transformation $a_i^{\alpha}\to a_i^{\alpha} q_i^{(m)}$,
$b_i^{\alpha}\to b_i^{\alpha} (q_i^{(m)})^{-1}$ with
$\displaystyle{q_i^{(m)}=\exp \left (\int^{t_m}\Lambda_i^{(m)}dt\right )}$
eliminates the terms with $\Lambda_i^{(m)}$ and so we can put $\Lambda_i^{(m)}=0$.
In this way we obtain the Hamiltonian equations of motion for spin variables
in the higher times:
\beq\label{d6}
\p_{t_m}a_i^{\alpha}=\frac{\p {\cal H}_m}{\p b_i^{\alpha}}, \qquad
\p_{t_m}b_i^{\alpha}=-\frac{\p {\cal H}_m}{\p a_i^{\alpha}}.
\eeq
with ${\cal H}_m$ given by (\ref{int1}). 

\section{Time discretization of the trigonometric Gibbons-Hermsen model}

Our strategy is to substitute the pole ansatz for the discrete time wave functions
\beq\label{disc1}
\Psi_{\alpha \beta}^p=\Bigl (1-\frac{z}{\mu}\Bigr )^p e^{xz}
\left (C_{\alpha \beta}+\sum_i \frac{2\gamma w_i^{1/2}(p)a_i^{\alpha}(p)
c_i^{\beta}(p)}{w-w_i(p)}\right ),
\eeq
\beq\label{disc2}
\Psi_{\alpha \beta}^{\dag p}=\Bigl (1-\frac{z}{\mu}\Bigr )^{-p} e^{-xz}
\left (C^{-1}_{\alpha \beta}+\sum_i \frac{2\gamma w_i^{1/2}(p)c_i^{*\alpha}(p)
b_i^{\beta}(p)
}{w-w_i(p)}\right )
\eeq
and $w_{\alpha \beta}^{(1)}$ (see (\ref{t5}))
into the linear problems (\ref{l1a}), (\ref{l2a}) and identify the coefficients
in front of the poles $(w-w_i(p))^{-2}$, $(w-w_i(p\pm 1))^{-1}$ and
$(w-w_i(p))^{-1}$. (Note that the
constant term $S_{\alpha \beta}$ in $w^{(1)}_{\alpha \beta}(p)$ cancels in the combination
$w^{(1)}_{\alpha \beta}(p+1)-w^{(1)}_{\alpha \beta}(p)$ because 
the shift $p\to p+1$ is equivalent to a shift of times and
$S_{\alpha \beta}$ 
does not depend on the times.) We begin with the linear problem 
(\ref{l1a}) for $\Psi$. From cancellation of different 
poles we have the following conditions:
\begin{itemize}
\item
At $\frac{1}{(w-w_i(p))^2}$: \phantom{a} $b_i^{\nu}(p)a_i^{\nu}(p)=1$;
\item
At $\frac{1}{w-w_i(p+1)}$:  $$
(z-\mu )c_i^{\beta}(p+1)=-w_i^{1/2}(p)
\tilde b_i^{\beta}(p+1)-2\gamma
\sum_j\frac{w_i^{1/2}(p)w_j^{1/2}(p)
b_i^{\nu}(p+1)a_j^{\nu}(p)c_j^{\beta}(p)}{w_i(p+1)
-w_j(p)};$$
\item
At $\frac{1}{w-w_i(p)}$: 
$$
(z-\mu -2\gamma )a_i^{\alpha}(p)c_i^{\beta}(p)+
w_i^{1/2}(p)a_i^{\alpha}(p)\tilde b_i^{\beta}(p)
$$
$$
-2\gamma 
\sum_j\frac{w_j(p+1)a_j^{\alpha}(p+1)b_j^{\nu}(p+1)a_i^{\nu}(p)c_i^{\beta}(p)}{w_i(p)
-w_j(p+1)}
$$
$$
+2\gamma \sum_{j\neq i}\frac{w_i^{1/2}(p)w_j^{1/2}(p)
a_i^{\alpha}(p)b_i^{\nu}(p)
a_j^{\nu}(p)c_j^{\beta}(p)}{w_i(p) -w_j(p)}
+2\gamma \sum_{j\neq i}\frac{w_j(p)a_j^{\alpha}(p)b_j^{\nu}(p)
a_i^{\nu}(p)c_i^{\beta}(p)}{w_i(p) -w_j(p)}=0.
$$
\end{itemize}
Introduce the matrices
\beq\label{disc3}
L_{ij}(p)=-\delta_{ij}\frac{\dot x_i(p)}{2}-2\gamma (1-\delta_{ij})\,
\frac{w_i^{1/2}(p)w_j^{1/2}(p)b_i^{\nu}(p)a_j^{\nu}(p)}{w_i(p)-w_j(p)}
\eeq
(the same Lax matrix as in (\ref{t8})) and
\beq\label{disc4}
M_{ij}(p)=2\gamma \, 
\frac{w_i^{1/2}(p+1)w_j^{1/2}(p)b_i^{\nu}(p+1)a_j^{\nu}(p)}{w_i(p+1)-w_j(p)},
\eeq
then the above conditions can be written as
\beq\label{disc5}
\left \{
\begin{array}{l}
\displaystyle{(z-\mu )c_i^{\beta}(p+1)=-w_i^{1/2}(p+1)\tilde b_i^{\beta}(p+1)-
\sum_j M_{ij}(p)c_j^{\beta}(p)}
\\ \\
\displaystyle{ a_i^{\alpha}(p)\underbrace{\left [\sum_j\Bigl ((z-\gamma)
\delta_{ij}-L_{ij}(p)\Bigr )
c_j^{\beta}(p)+w_i^{1/2}(p)\tilde b_i^{\beta}(p)\right ]}_{=0}}
\\ \\
\displaystyle{\phantom{aaaaaaa}
+c_i^{\beta}(p)\left [ \sum_j a_j^{\alpha}(p+1)(W^{1/2}(p+1)
M(p)W^{-1/2}(p))_{ji}\right. }
\\ \\
\displaystyle{\phantom{aaaaaaaaaaaaa}
\left. +\sum_j a_j^{\alpha}(p)(W^{1/2}(p)L(p)W^{-1/2}(p))_{ji}
-(\mu +\gamma ) a_i^{\alpha}(p)\right ]=0.}
\end{array}
\right.
\eeq
The first line in the second equation vanishes by virtue of (\ref{t6}). 
Therefore, we have the following equations:
\beq\label{disc6}
(z-\mu ){\bf c}^{\beta}(p+1)=-W^{1/2}(p+1)\tilde {\bf b}^{\beta}(p+1)-
M(p){\bf c}^{\beta}(p),
\eeq
\beq\label{disc7}
{\bf a}^{\alpha T}(p+1)W^{1/2}(p+1)M(p)W^{-1/2}(p)+
{\bf a}^{\alpha T}(p)W^{1/2}(p)L(p)W^{-1/2}(p)=(\mu +\gamma ){\bf a}^{\alpha T}(p).
\eeq
A similar solution of the linear problem (\ref{l2a}) for $\Psi^{\dag}$ gives the
equations
\beq\label{disc8}
(z-\mu ){\bf c}^{*\alpha}(p-1)=\tilde {\bf a}^{\alpha T}(p-1)W^{1/2}(p-1)-
{\bf c}^{*\alpha}(p)M(p-1),
\eeq
\beq\label{disc9}
W^{-1/2}(p)M(p-1)W^{1/2}(p-1){\bf b}^{\beta }(p-1)+
W^{-1/2}(p)L(p)W^{1/2}(p){\bf b}^{\beta}(p)=(\mu -\gamma ){\bf b}^{\beta}(p).
\eeq

A simple calculation, similar to the one done in \cite{Z19aa}, shows that the
compatibility condition of equations (\ref{t6}), (\ref{disc6}) is 
the discrete Lax equation
\beq\label{disc10}
L(p+1)M(p)=M(p)L(p)
\eeq
which holds true provided equations (\ref{disc7}), (\ref{disc9}) are satisfied.

Equations (\ref{disc7}), (\ref{disc9}) are equations of motion of the discrete time
trigonometric Gibbons-Hermsen model. Let us consider equation (\ref{disc7}) and represent it
in a somewhat better form. In order to do this, write it in the form
$$
2\gamma \sum_k \frac{w_i(p)a_k^{\alpha}(p+1)b_k^{\nu}(p+1)a_i^{\nu}(p)}{w_k(p+1)-w_i(p)}
+2\gamma \sum_{k\neq i}
\frac{w_i(p)a_k^{\alpha}(p)b_k^{\nu}(p)a_i^{\nu}(p)}{w_i(p)-w_k(p)}
$$
$$
+2\gamma \sum_k a_k^{\alpha}(p+1)b_k^{\nu}(p+1)a_i^{\nu}(p)-2\gamma
\sum_{k\neq i}a_k^{\alpha}(p)b_k^{\nu}(p)a_i^{\nu}(p)
-(\mu +\gamma )a_i^{\alpha}(p)-\frac{\dot x_i(p)}{2}\, a_i^{\alpha}(p)=0
$$
and add it to the original equation taking into account that
$$
\sum_k a_k^{\alpha}(p+1)b_k^{\nu}(p+1)=\sum_k a_k^{\alpha}(p)b_k^{\nu}(p).
$$
This follows from the fact that 
$\displaystyle{
\sum_i a_i^{\alpha}b_i^{\beta}}$ is an integral of motion, i.e., 
$\displaystyle{
\p_{t_m}\! \Bigl (\sum_i a_i^{\alpha}b_i^{\beta}\Bigr )=0}$ for all $m$. Indeed, we have
$$
\p_{t_m}\! \Bigl (\sum_i a_i^{\alpha}b_i^{\beta}\Bigr )=\sum_i
\left (b_i^{\beta}\frac{\p {\cal H}_m}{\p b_i^{\alpha}}-
a_i^{\alpha}\frac{\p {\cal H}_m}{\p a_i^{\beta}}\right )
$$
and this is zero because ${\cal H}_m$ is a linear combination of 
$H_k=\mbox{tr}\, L^k$ and
$$
\sum_i \left (b_i^{\beta}\mbox{tr} \, \Bigl (\frac{\p L}{\p b_i^{\alpha}}\, L^{m-1}\Bigr )-
a_i^{\alpha}\mbox{tr}\, \Bigl (\frac{\p L}{\p a_i^{\beta}}\, L^{m-1}\Bigr )\right )
$$
$$
=\sum_i \sum_{j,k}
\left (b_i^{\beta} \frac{\p L_{jk}}{\p b_i^{\alpha}}\, L^{m-1}_{kj}-
a_i^{\alpha}\frac{\p L_{jk}}{\p a_i^{\beta}}\, L^{m-1}_{kj}\right )
$$
$$
=2\gamma \sum_i \sum_{j\neq k}(\delta_{ik}-\delta_{ij})\,
\frac{w_j^{1/2}w_k^{1/2}b_j^{\beta}a_k^{\alpha}}{w_j-w_k}\, L^{m-1}_{kj}=0.
$$
As a result, we obtain the equation
\beq\label{disc11}
\begin{array}{c}
\displaystyle{
\gamma \! \sum_k \coth (\gamma (x_k (p+1)-x_i(p))a_k^{\alpha}(p+1)
b_k^{\nu} (p+1)a_i^{\nu}(p)}
\\ \\
\displaystyle{
=\gamma \! \sum_{k\neq i}
\coth (\gamma (x_k (p)-x_i(p))a_k^{\alpha}(p)
b_k^{\nu} (p)a_i^{\nu}(p)
+\frac{\dot x_i(p)}{2}\, a_i^{\alpha}(p)+\mu a_i^{\alpha}(p)}.
\end{array}
\eeq
A similar transformation of equation (\ref{disc9}) leads to the equation
\beq\label{disc12}
\begin{array}{c}
\displaystyle{
\gamma \! \sum_k \coth (\gamma (x_i (p)-x_k(p-1))b_k^{\alpha}(p-1)
b_i^{\nu} (p)a_k^{\nu}(p-1)}
\\ \\
\displaystyle{
=\gamma \! \sum_{k\neq i}
\coth (\gamma (x_i (p)-x_k(p))b_k^{\alpha}(p)
b_i^{\nu} (p)a_k^{\nu}(p)
+\frac{\dot x_i(p)}{2}\, b_i^{\alpha}(p)+\mu b_i^{\alpha}(p)}.
\end{array}
\eeq
Multiply the first equation by $b_i^{\alpha}(p)$ and sum over $\alpha$, then 
multiply the second equation by $a_i^{\alpha}(p)$, sum over $\alpha$ and take into 
account the constraint $b_i^{\nu}a_i^{\nu}=1$.
Subtracting the resulting equations, we eliminate $\dot x_i (p)$ and obtain the 
equations of motion (\ref{I2}):
\beq\label{disc13}
\begin{array}{c}
\displaystyle{
\sum_j 
\coth (\gamma (x_i(p)-x_j(p+1))
b_i^{\nu}(p)a_j^{\nu}(p+1))b^{\beta}_j(p+1)
a_i^{\beta}(p)}
\\ \\
\displaystyle{
+\sum_j \coth (\gamma (x_i(p)-x_j(p-1))
b_i^{\nu}(p)a_j^{\nu}(p-1))b^{\beta}_j(p-1)
a_i^{\beta}(p)}
\\ \\
\displaystyle{=
2\sum_{j\neq i} \coth (\gamma (x_i(p)-x_j(p))
b_i^{\nu}(p)a_j^{\nu}(p))b^{\beta}_j(p)
a_i^{\beta}(p)}.
\end{array}
\eeq
These equations of motion generalize the ones for the rational Gibbons-Hermsen
model obtained in \cite{Z19aa}. They look like the Bethe ansatz equations for the
quantum trigonometric Gaudin model ``dressed'' by the spin variables. 
In the continuum limit the equations of motion (\ref{t14}) are reproduced.

\section*{Acknowledgments}

The work of A.Z. was supported by the Russian Science Foundation under grant 19-11-00062.


\end{document}